# Pressure-induced dimensional crossover in a kagome superconductor

Fanghang Yu,[1, §] Xudong Zhu,[1, §] Xikai Wen,[1] Zhigang Gui,[1] Zeyu Li,[1] Yulei Han,[4] Tao Wu,[1] Zhenyu Wang,[1] Ziji Xiang,[1] Zhenhua Qiao,[1*] Jianjun Ying,[1†] and Xianhui Chen[1,2,3,‡]

[1]Hefei National Laboratory for Physical Sciences at Microscale and Department of Physics, and CAS Key Laboratory of Strongly-coupled Quantum Matter Physics, University of Science and Technology of China, Hefei, Anhui 230026, China

[2]CAS Center for Excellence in Quantum Information and Quantum Physics, Hefei, Anhui 230026, China

[3]Collaborative Innovation Center of Advanced Microstructures, Nanjing 210093, People's Republic of China

[4]Department of Physics, Fuzhou University, Fuzhou, Fujian 350108, China

*E-mail: qiao@ustc.edu.cn
†E-mail: yingjj@ustc.edu.cn
‡E-mail: chenxh@ustc.edu.cn
§ These authors contributed equally to this work

**The recently discovered kagome superconductors $AV_3Sb_5$ exhibit tantalizing high-pressure phase diagrams, in which a new dome-like superconducting phase emerges under moderate pressure. However, its origin is as yet unknown. Here, we carried out the high-pressure electrical measurements up to 150 GPa, together with the high-pressure X-ray diffraction measurements and first-principles calculations on $CsV_3Sb_5$. We find the new superconducting phase to be rather robust and inherently linked to the interlayer Sb2-Sb2 interactions. The formation of Sb2-Sb2 bonds at high pressure tunes the system from two-dimensional to three-dimensional and pushes the $P_z$ orbital of Sb2 upward across the Fermi level, resulting in enhanced density of states and increase of $T_C$. Our work demonstrates that the dimensional crossover at high pressure can induce a topological phase transition and is related to the abnormal high-pressure $T_C$ evolution. Our findings should apply for other layered materials.**

Due to the unusual lattice geometry, the materials with kagome lattice provide a fertile playground to study the frustrated, novel correlated and topological electronic states. Recently, a new family of quasi two-dimensional kagome metals $AV_3Sb_5$ (*A*=K, Rb, Cs) has attracted tremendous attention [1]. These materials crystallize in the *P*6/*mmm* space group, forming layers of ideal kagome nets of V ions coordinated by Sb. Besides topological properties, $AV_3Sb_5$ exhibits intriguing physical properties, including charge density wave (CDW) [1,2] and superconductivity [3-5]. Intertwining the superconductivity and CDW state by involving nontrivial topology of band structures results in many exotic electronic properties in this type of materials [2,6-16].

High pressure is a clean method to tune the electronic properties without introducing any impurities. Therefore, pressure is often used as a control parameter to tune topological properties, superconducting and CDW transition temperatures. Besides the exotic ambient-pressure properties, the superconductivity in compressed $CsV_3Sb_5$ shows abnormal behavior. Recent high-pressure works indicate unconventional competition between CDW and superconductivity in $CsV_3Sb_5$. In contrast to other conventional CDW superconductors, a two-dome superconductivity behavior is observed before CDW is completely suppressed [9,17]. Further elevating the pressure, the original superconducting phase (SC I) can be gradually suppressed. More intriguingly, a new dome-like superconducting phase (SC II) emerges above 15 GPa [18-20]. Such new dome-like superconducting phase is also observed in compressed $KV_3Sb_5$ and $RbV_3Sb_5$ single crystals [21-23]. Despite that the rich high-pressure phase diagram is discovered in experiments, its origin is still not understood. The van Hove singularities move away from the Fermi level under pressure [24], which is crucial for the emergence of different Fermi surface instabilities in the $AV_3Sb_5$ family. Recent theoretical works proposed that the magnetism of V may greatly suppress $T_C$ in the low-pressure region[25]. The anisotropic compression is also considered to play an important role during the compression process[26]. In order to figure out the mechanism of the abnormal high-pressure phase diagram in $CsV_3Sb_5$, detailed experimental and theoretical explorations are highly desired.

By applying pressure, the 2-dimensional (2D) system can be gradually tuned to a 3D system due to the rapid reduction of interlayer distance. Therefore, it would be interesting to check whether superconductivity is tied to the dimensional crossover in $CsV_3Sb_5$. The angle-resolved photoemission spectroscopy (ARPES) and theoretical calculations demonstrated the existence of $\mathbb{Z}_2$ nontrivial topological band structure in $CsV_3Sb_5$ at ambient pressure[3]. It is also interesting to check how the topological band structure evolves under pressure. Here, by combing the high-pressure electrical transport, X-ray diffraction (XRD) and first principles calculations, we are able to extent its high-pressure phase diagram to 150 GPa. In contrast to previous results, the superconductivity persists in the whole pressure range. After CDW is completely suppressed, the $T_C$ is rapidly suppressed to a minimum of 1 K at around 14 GPa, then $T_C$ can be enhanced to 5.8 K at ~45 GPa. With further raising the pressure, $T_C$ starts to decrease slowly and can be suppressed to 3.2 K at 150 GPa. Our high-pressure XRD measurements suggest the intimate relation between $T_C$ and $c/a$ value. Combing with the first principles calculations, our results indicate that the 2D $CsV_3Sb_5$ would gradually evolve to 3D structure at high pressure, in which the interlayer Sb2-Sb2 bonds form as $CsSb_4$ building blocks. Such 2D-to-3D crossover would dramatically modify the band structure, which could be beneficial for the superconductivity and can alter the topological properties.

Single crystals of $CsV_3Sb_5$ were synthesized via a self-flux growth method, similar to the previous reports[11]. Diamond anvils with various culets (100 to 300 µm) were used for high-pressure transport measurements. NaCl was used as a pressure transmitting medium and the pressure was calibrated by using the shift of ruby florescence and diamond anvil Raman at room temperature. During transport measurements, the pressure was applied at room temperature using the miniature diamond anvil cell. The transport measurements were carried out using the Quantum Design PPMS9 or a $^3$He cryostat (HelioxVT, Oxford Instruments). The high-pressure synchrotron XRD was carried out at room temperature at the beamline BL15U1 of the Shanghai Synchrotron Radiation Facility (SSRF) with a wavelength of $\lambda = 0.6199$ Å. A symmetric diamond anvil cell with a pair of 200 µm culet size anvils was used to generate pressure. 70 µm sample chamber is drilled from the Re gasket and Daphne 7373 oil was loaded as a pressure transmitting medium.

For theoretical analysis, the structural optimization and electronic structure calculations were performed by using density functional theory with projected augmented-wave method [27] as implemented in the Vienna ab initio Simulation Package (VASP) package [28-30]. For exchange correlation, we used the generalized gradient approximation of the Perdew-Burke-Ernzerhof [31] functional with DFT-D3 van der Waals correction [32], the spin-orbit coupling (SOC) is considered in our calculation. The plane wave energy cutoff was set to be 600 eV with an energy precision of

$10^{-6}$ eV. Both lattice parameters and atomic positions were fully optimized under a series of pressures until the Hellmann-Feynman forces on each ion were less than $10^{-3}$ eV/Å. At ambient pressure, the optimized lattice constant of CsV$_3$Sb$_5$ is consistent with the experimental value [3].

Phonon dispersion calculations were performed by using a $3 \times 3 \times 2$ supercell via the finite displacement method as implemented in the PHONOPY code [33]. Numbering of the irreducible representations (see Fig. S8-S11) refers to the notation of Koster *et al* [34]. $\mathbb{Z}_2$ topological invariant calculations were performed using the Irvsp [35] program in conjunction with VASP.

We performed resistivity measurements on CsV$_3$Sb$_5$ single crystals under pressure to track the evolution of superconductivity. Temperature dependence of resistivity for CsV$_3$Sb$_5$ under various pressures is shown in Fig. 1. After CDW is completely suppressed above 2 GPa, the maximum $T_C$ around 8 K can be achieved. By increasing the pressure, the superconducting transition of CsV$_3$Sb$_5$ can be gradually suppressed to 1 K around 14 GPa as shown in Fig. 1(a) and (b). Further increasing the pressure, $T_C$ starts to increase and reaches 5.8 K around 45 GPa as shown in Figs. 1(b) and (c). Then, $T_C$ gradually decreases with increasing the pressure. Surprisingly, we can still observe superconductivity with $T_C$ around 3.2 K even at 150 GPa. We can map out the high-pressure phase diagram as shown in Fig. 2(a). Our work demonstrates that the previous reported SC I and SC II phases are indeed connected with each other. The maximum $T_C$ observed in the SC II is slightly higher than the previous results, such difference may arise from the slightly different hydrostatic conditions in the high-pressure experiments. The SC II phase is rather robust, which can persist up to at least 150 GPa. Meanwhile, the magnetoresistance (MR) gradually decreases with increasing pressure as shown in Fig. 2(c) in the supplementary materials. The MR shows a sudden reduction around 14 GPa, coincident with the $T_C$ minimum. Above 45 GPa, both $T_c$ and MR becomes weakly pressure-dependent. All these results directly indicate that MR is closely related to $T_C$.

To clarify the physical origin of the unusual $T_C$ evolution under pressure, we performed high-pressure XRD measurements as shown in Fig. 3. The XRD patterns collected at different pressures are displayed in Fig. 3(a). We performed Le Bail fit for all the XRD patterns. All the XRD patterns can be well fitted using its ambient pressure structure, and selected fitting results are shown in Fig.S4 in the supplementary materials. However, we should point out that the linewidth of the diffraction peaks significantly increases due to the pressure inhomogeneity. We can extract the linewidth of each main diffraction peaks as shown in Fig.S5 in the supplementary materials. Since the Rietveld refinement was not possible due to the pressure inhomogeneity for our XRD patterns, our preliminary analysis with the Le Bail model shows that the lattice symmetry of CsV$_3$Sb$_5$ does not change as a function of pressure. However, the underlying crystal structure at high pressure can not be settled due to the limitation of the Le Bail fit. Therefore, the intriguing $T_C$ behavior under pressure cannot attribute to a sudden lattice symmetry change. We can extract the lattice parameter *a* and *c* as a function of pressure shown in Fig. 3(b). The derived cell volume is shown in Fig. 3(c), which can be well fitted by using the Birch–Murnaghan equation of state with the derived bulk modulus $B_0 = 37.7$ GPa. We note that the lattice parameter *c* decreases rather rapidly at lower pressure with two slope changes observed at 15 and 50 GPa, respectively. These results indicate that the interlayer interaction becomes more pronounced at high pressure.

CsV$_3$Sb$_5$ crystallizes in the *P*6/*mmm* space group. The vanadium sublattice is a structurally perfect kagome lattice. The Sb1 sublattice is a simple hexagonal net, centered on each kagome hexagon. Between different kagome V$_3$Sb layers, the Sb2 sublattice forms two honeycomb-type Sb sheets

(antimonene) embedded by Cs atom as shown in Fig. 4(a). By decreasing the *c* lattice parameter, we find that the interlayer coupling of different Sb2 sublattice is gradually enhanced and finally forms Sb2-Sb2 bonds at high pressure. To investigate the dynamical structural stability at high pressure, we calculated the phonon dispersion spectra from 5 to 60 GPa (see Fig. S6). The absence of imaginary frequencies above 5 GPa indicates the structural stability of $CsV_3Sb_5$ under pressure.

To figure out the origin of SC II phase, we calculated the lattice parameters and electronic properties at various pressures. The optimized lattice parameters *a*, *c* and the distance between different Sb2 atoms are plotted in Fig. S7. The calculated axial ratio *c/a* is consistent with the experimental results below 30 GPa as plotted in Fig. 4(b) and the difference between *c/a* results in calculation and experiment may arise from the vdW correction in calculation at higher pressure. One can find that the *c/a* ratio rapidly decreases at low pressure, and the decay rate of *c/a* exhibits a crossover behavior around 15 GPa, which is consistent with the $T_C$ minimum from the resistivity measurements. By applying pressure, the honeycomb-like Sb2 sheets between kagome $V_3Sb$ layers get close to each other along the *c* axis. The distance of Sb2 sheets along *c* axis is compressed to be less than 3.2 Å above 15 GPa. Under compressive stress, the formation of the Sb2-Sb2 bonds between different kagome $V_3Sb$ layers determines the structural evolution of $CsV_3Sb_5$. Therefore, as shown in Fig. 4(c)-(f), the $p_z$ orbital of Sb2 gradually dominates near the Fermi energy $E_F$, comparing with that of Sb1 above 15 GPa.

When the pressure exceeds 45 GPa, which marks the $T_C$ maximum in the SC II phase, the *c/a* ratio becomes almost a constant and Sb2 atoms get closer to each other along *c* axis than that in the honeycomb sublattice plane. The rapid reduction of *c/a* below 50 GPa indicates the quasi-2D structure can be tuned to 3D form under pressure. The nearly constant *c/a* value above 50 GPa represents the isotropic compression ratio at high pressure, indicating the system becomes 3D. The structural evolution from 2D to 3D nature can be attributed to the formation of the Sb2-Sb2 bonding between different $V_3Sb$ layers as shown in Fig. 4(a). Such phenomenon is similar to the collapsed tetragonal structural phase transition in 122 iron-pnictide superconductors [36]. The formation of Sb2-Sb2 bonds can drastically affect the band structure, making the $p_z$ orbital of Sb2 move upward quickly at the high symmetry point *A*, which can then enhance the density of states near $E_F$ and support the enhancement of $T_C$ above 14 GPa in the experiments.

We also investigated the evolution of topological properties of $CsV_3Sb_5$ under high pressure. The structure of $CsV_3Sb_5$ displays the dynamical stability under pressures (see Fig. S6), and it also possesses both time-reversal and inversion symmetries under pressures. We calculated the $\mathbb{Z}_2$ topological invariant at various pressures between the bands crossing $E_F$ by analyzing the parity of the wave function at the time-reversal invariant momentum (TRIM) points [37], which can be computed by using:

$$(-1)^\nu = \prod_{i=1}^{4} \delta_i$$

Where *v* is the $\mathbb{Z}_2$ topological invariant, and $\delta_i$ is the parity product of bands at the TRIM point *i* (Γ, *A*, *L*, or *M*).

Figure 5 shows the results of $\mathbb{Z}_2$ topological invariant calculation. Given that there is no clear division between conduction and valence states in this kind of kagome metals, we used the band indexes from DFT calculations to distinguish those bands crossing the $E_F$. We can see from Fig. 5(a) that the three bands below $E_F$ display topological nontrivial characters below 20 GPa. With increasing the pressure, the #75 band keeps $\mathbb{Z}_2 = 1$ whereas #71 band occurs a topological phase transition above 40 GPa. Interestingly, the #73 band displays a topological transition at 19-20 GPa as shown in Fig. 5(b), which is close to the superconductivity phase boundary. The topological phase transition of band #73 can be attributed to the band gap closing at Γ point between bands #73 and #75 (highlighted with orange circles), which induces the parity change of $\delta_\Gamma$, see Fig. 5(e) and Fig. S8-S12. In addition, the magnetoresistance also presents a pressure-induced change which is coincident with the topological phase transition around 20 GPa (see Fig. 2(c)), possibly originating from the reduction of carrier mobility [38,39].

The superconductivity in $CsV_3Sb_5$ is highly tunable and rather sensitive to the pressure, which can be attributed to the enhancement of interlayer coupling. The interlayer coupling in $CsV_3Sb_5$ is weak at ambient pressure, thus the samples are quasi-2D and can be easily mechanically cleaved [40]. The pressure can dramatically change the lattice parameters, especially the interlayer distance that would greatly modify the band structure. The $p_z$ orbital of Sb2 gradually crosses the Fermi surface above 10 GPa and enhances the density of states. This can explain the enhancement of $T_C$ above 14 GPa in our experiments. Above 45 GPa, the sample become isotropic, and $T_C$ shows weak pressure dependence. The dramatic change of band structure can also alter the topological properties. A pressure-induced topological phase transition is observed above 20 GPa. The pressure induced new dome-like superconductivity is universal in $AV_3Sb_5$ family. The maximum $T_C$ and pressure range in SC II increases with increasing the ionic radius of $A$. The smaller radius of $A$ provides additional chemical pressure that would significantly lower the 2D-3D crossover pressure. However, it is still unclear how the radius of $A$ affects the maximum $T_C$ in the $AV_3Sb_5$ family.

In conclusion, we found that the two superconducting phases SC I and SC II in $CsV_3Sb_5$ are actually connected with each other. The SC II is rather robust and can persist up to at least 150 GPa. The abnormal evolution of superconductivity can be attributed to the formation of interlayer Sb2-Sb2 bonding that enhances the system's three-dimensionality at high pressure. In addition, we also found a pressure-induced topological phase transition at 20 GPa.

**Acknowledgements**

This work was supported by the Anhui Initiative in Quantum Information Technologies (Grant No. AHY160000), the National Key Research and Development Program of the Ministry of Science and Technology of China (Grants No. 2019YFA0704900 and No. 2017YFA0303001), the Science Challenge Project of China (Grant No. TZ2016004), the Key Research Program of Frontier Sciences, CAS, China (Grant No. QYZDYSSWSLH021), the Strategic Priority Research Program of the Chinese Academy of Sciences (Grant No. XDB25000000), the National Natural Science Foundation of China (Grants No. 11888101, No. 11534010, No. 11974327 and No. 12004369), the Collaborative Innovation Program of Hefei Science Center, CAS, (Grant No. 2020HSC-CIP014) and the Fundamental Research Funds for the Central Universities (Grants No. WK3510000011, No. WK3510000010, and No. WK2030020032). High-pressure synchrotron XRD work was performed at the BL15U1 beamline, Shanghai Synchrotron Radiation Facility (SSRF) in China. We are grateful to AM-HPC and the Supercomputing Center of USTC for providing high-performance computing resources.

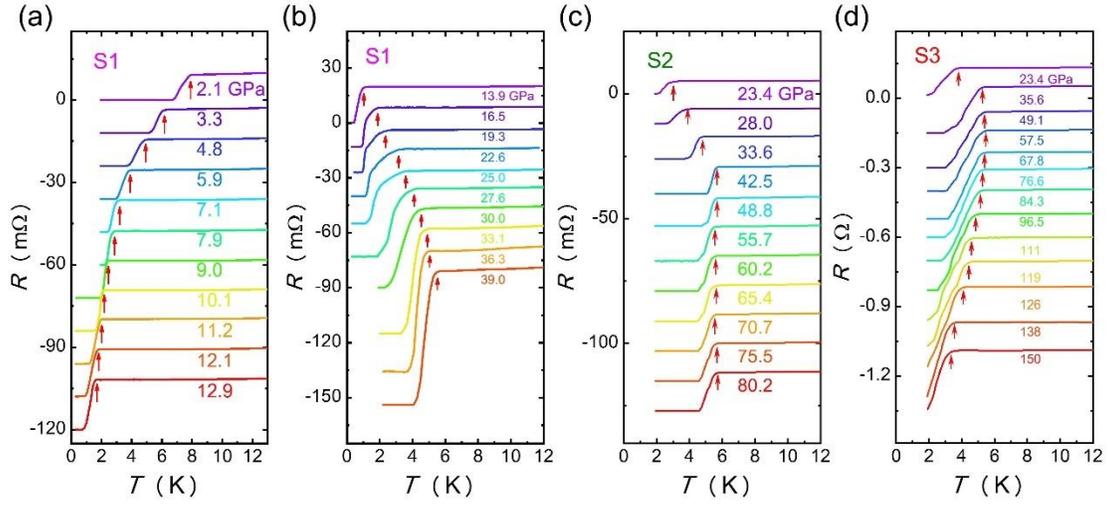

Figure 1. Temperature dependence of resistivity in $CsV_3Sb_5$ single crystals under high pressure. (a) $T_C$ gradually suppressed with increasing the pressure for sample 1 below 14 GPa. (b) $T_C$ start to increase with increasing the pressure above 14 GPa. (c) $T_C$ reaches its maximum value around 45 GPa for sample 2 in SC II phase. (d) $T_C$ gradually suppressed with increasing the pressure above 45 GPa. At 150 GPa, $T_C$ can be suppressed to 3.2 K. All the curves were shifted vertically for clarity.

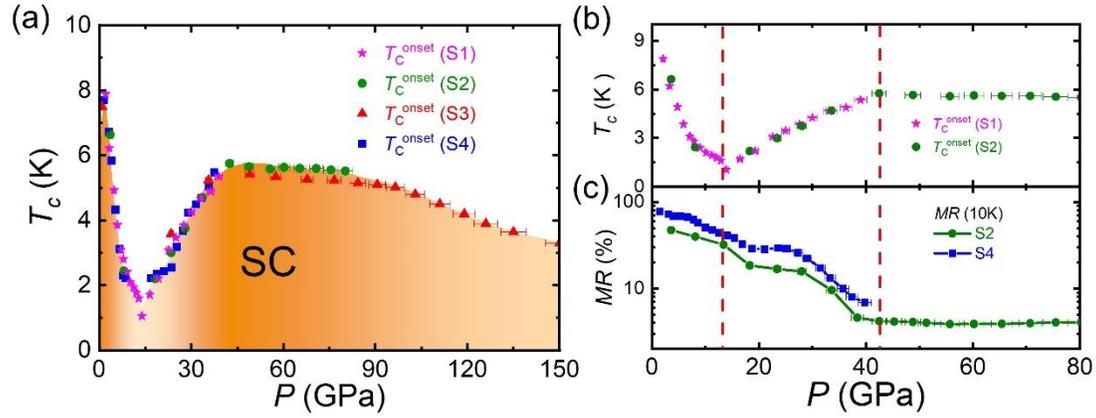

Figure 2. (a) Temperature-pressure phase diagram for $CsV_3Sb_5$ single crystal. The superconducting transition temperatures were measured on various samples. A new robust dome-like superconducting phase emerges above 14 GPa and can persists to the highest pressure we have measured (150 GPa). (b) Pressure dependence of superconducting transition temperatures. (c) The magnetoresistance (9T and 10 K) evolution under pressure.

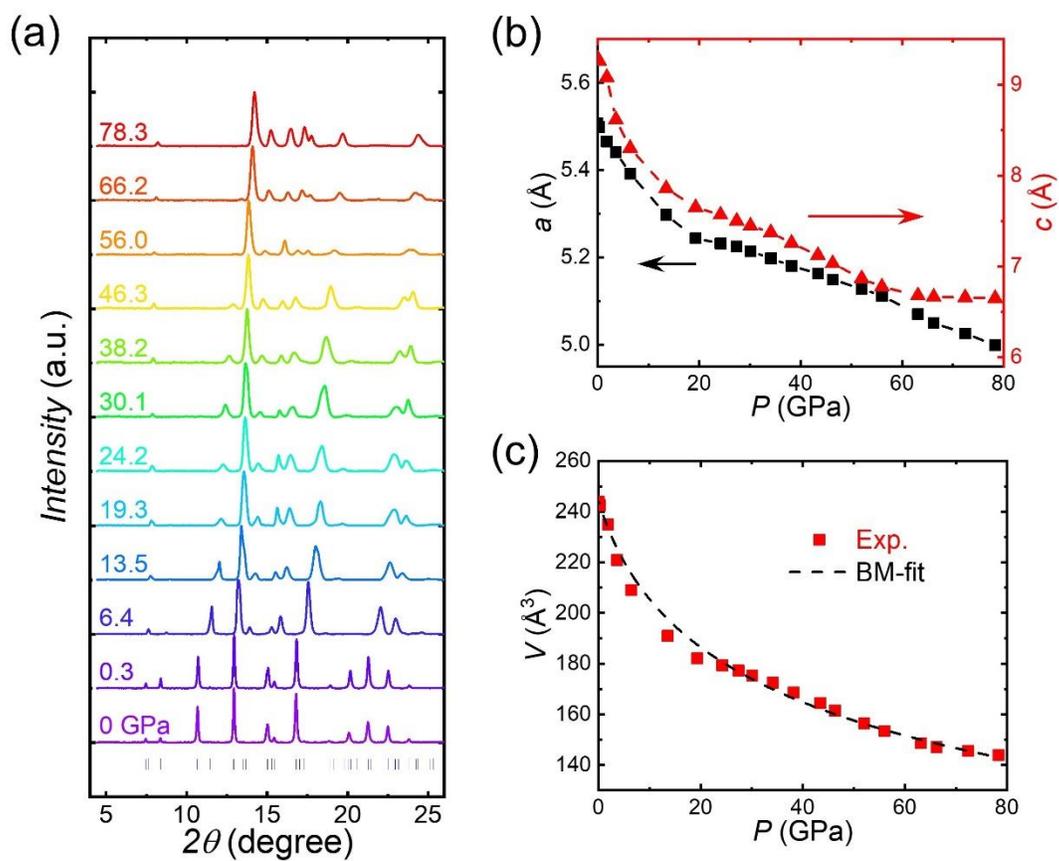

Figure 3. (a) XRD patterns of $CsV_3Sb_5$ under high pressure up to 78.3 GPa with an incident wavelength $\lambda = 0.6199$ Å. (b) Pressure dependence of the lattice parameters $a$ and $c$ for $CsV_3Sb_5$. (c) The derived cell volume as a function of pressure for $CsV_3Sb_5$. The black dashed line is a fitted curve by the Birch-Murnaghan equation of state with the derived bulk modulus $B_0 = 37.7$ GPa.

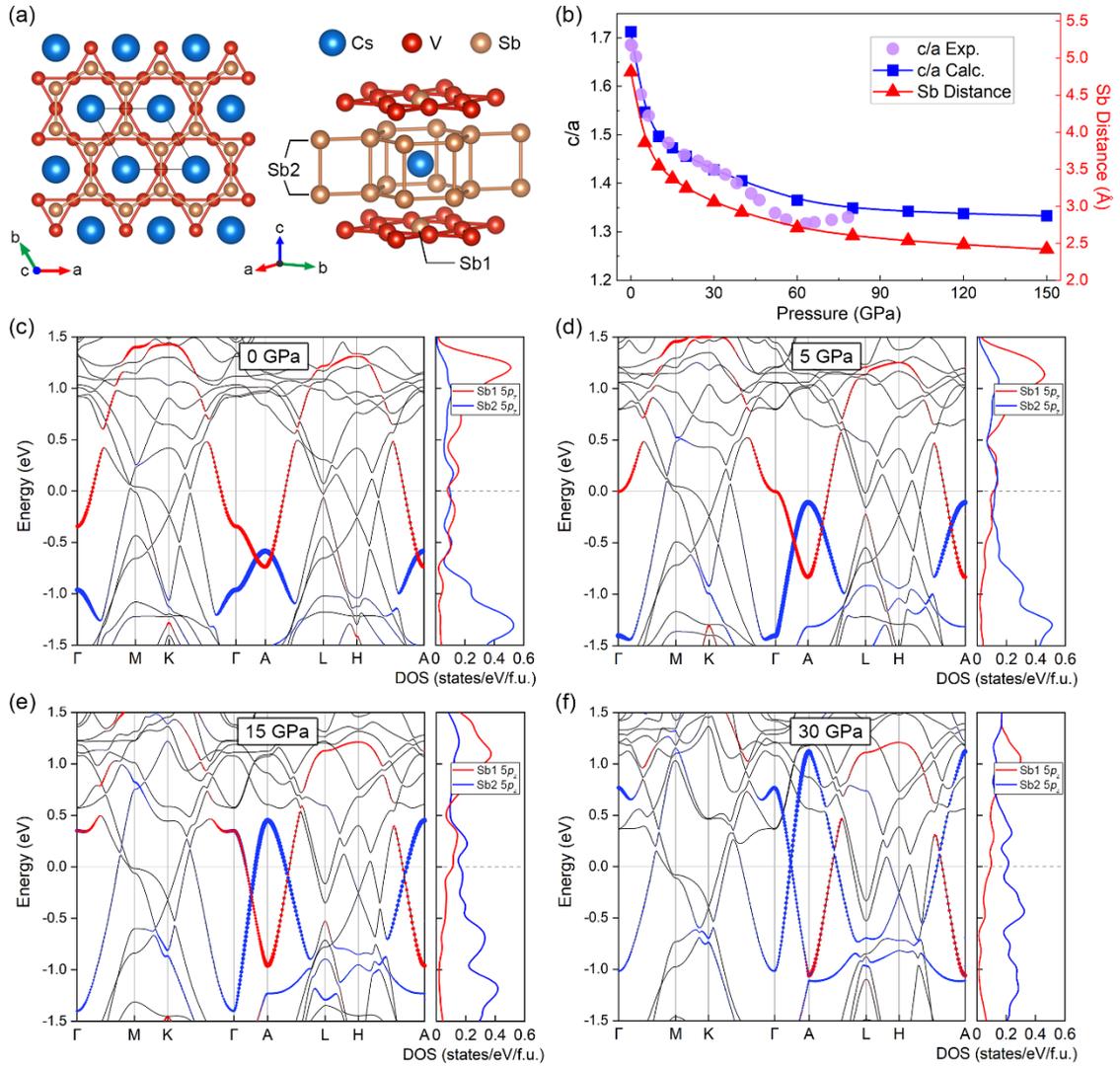

Figure 4. (a) The top and side view of $CsV_3Sb_5$ crystal structure. The Sb layers between different kagome $V_3Sb$ layers appeal to each other along *c* axis under compressive stress. (b) Experimental and calculated *c/a* ratio and the calculated Sb2-Sb2 bond length as a function of pressure. The projected band structures (PBAND) and partial density of states (PDOS) of Sb1 $p_z$ and Sb2 $p_z$ orbitals at (c) 0 GPa, (d) 5 GPa, (e) 15 GPa and (f) 30 GPa, respectively.

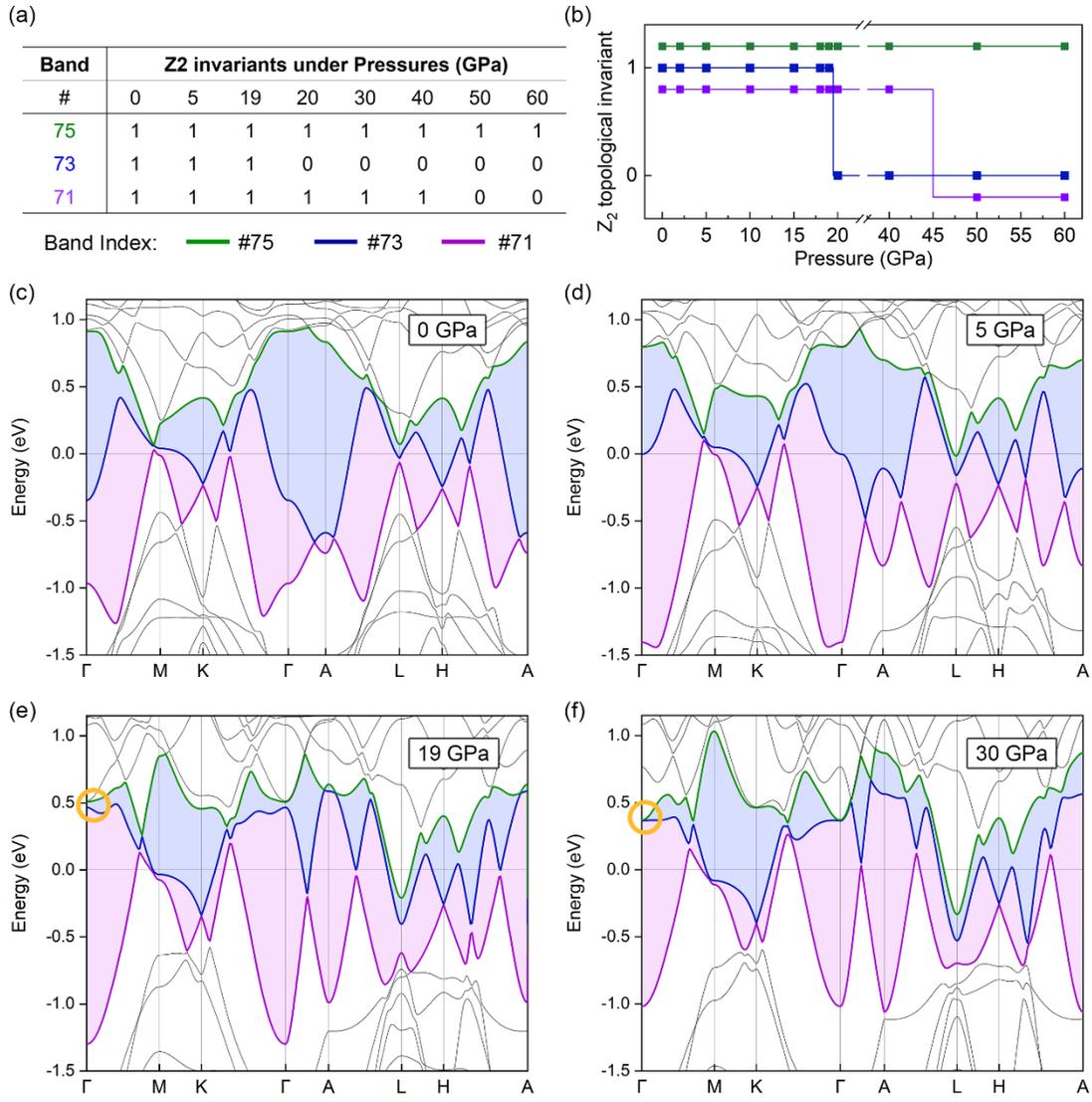

Figure 5. (a) $\mathbb{Z}_2$ topological invariants of the three bands across the Fermi level under different pressures. (b) The evolution of $\mathbb{Z}_2$ topological invariants of the three bands under pressures. The curves were shifted vertically for clarity. (c)-(f): The band structures of CsV$_3$Sb$_5$ at (c) 0 GPa, (d) 5 GPa, (e) 19 GPa and (f) 30 GPa, respectively. The three doubly degenerate bands across the Fermi level are labeled with green (#75), blue (#73), and purple (#71), respectively.